\documentstyle[epsfig,psfig,rotate]{aa}
\def\simlt{\lower.5ex\hbox{$\; \buildrel < \over \sim \;$}}

\begin{document}
\thesaurus{3 (11.04.1;              
              11.12.2;              
              12.03.3;              
              12.05.1;              
              12.12.1;              
              13.18.1)              
}

\authorrunning{F. Bertoldi et al.}
\titlerunning{Three high-redshift millimeter sources}

\title{Three high-redshift millimeter sources and their 
radio and near-infrared identifications}

\author{
F. Bertoldi\inst{1}\and    
C. L. Carilli\inst{1,}\inst{2}  \and 
K. M. Menten\inst{1}  \and      
F. Owen\inst{2,6}    \and 
A. Dey\inst{3}     \and 
F. Gueth\inst{1}   \and 
J. R. Graham\inst{4}  \and
E. Kreysa\inst{1}  \and
M. Ledlow\inst{5,6}   \and
M. C. Liu\inst{4}  \and 
F. Motte\inst{1}   \and 
L. Reichertz\inst{1} \and
P. Schilke\inst{1} \and
R. Zylka\inst{1}   
}

\institute{
Max-Planck-Institut f\"ur Radioastronomie,
Auf dem H\"ugel 69, D-53121 Bonn, Germany
\and
National Radio Astronomy Observatory, P.O. Box O,
Socorro, NM 87801, USA
\and
National Optical Astronomy Observatory, Tucson, AZ 85719, USA
\and
University of California at Berkeley, Astronomy Department,
Berkeley, CA 94720, USA
\and
Gemini Observatory, AURA, Casilla 603, La Serena, Chile, S.A.
\and
Kitt Peak National Observatory, NOAO, Tucson, AZ 85719, USA
}

\offprints{Bertoldi@MPIfR-Bonn.MPG.de}
\date{Received 14 March 2000; accepted }
\maketitle

\begin{abstract}
  
  We present millimeter wavelength detections of three faint sources
  that are most likely high-redshift starburst galaxies. For one of the
  sources, which was previously discovered with SCUBA at 850\,$\mu$m, we
  present a detection with the IRAM interferometer at 240\,GHz (1.25 mm)
  that shows the object unresolved at an angular resolution of
  $2.\!''5$, and coincident within 1$''$ with a radio source and a
  galaxy detected in the near-infrared.  The two other sources were
  discovered in a deep 250\,GHz (1.2 mm) survey with the Max-Planck
  Millimeter Bolometer (MAMBO) array at the IRAM 30~m telescope.  Both
  have fluxes of $\approx 4$\,mJy and radio counterparts with a 1.4\,GHz
  flux density of $\approx 75~\mu$Jy.  Their radio--to--mm flux ratios
  suggest redshifts larger than 2.  Both sources are faint in the
  optical and near-infrared, one showing a 20.5\,mag $K$-band
  counterpart. From our data and that available in the literature, we
  estimate the redshift distribution of twenty-two faint mm and sub-mm
  sources and conclude that the majority of them are likely to be at $z
  > 2$.
  

\end{abstract}


\section{Introduction}
\label{se:introduction}

Recent determinations of the sky surface density of faint sub-millimeter
sources have revolutionized our understanding of the star formation
history of the universe by detecting a significant population of
dust-obscured, massive star forming galaxies at high redshift.
Detections at 350\,GHz of apparently high-redshift sub-millimeter
(sub-mm) sources in the Hubble Deep Field (Hughes et
al.~\cite{hughes98}), in three fields observed as part of the Canada-UK
Deep Submillimeter Survey (Lilly et al.~\cite{lilly99}; Eales et
al.~\cite{eales99}), and in a survey of galaxy clusters (Smail et
al.~\cite{smail97}; Ivison et al.~\cite{ivison98a}, \cite{ivison98b})
imply that optical studies may have under-estimated the integrated
cosmic star formation rate at $z\ge 2$ by at least a factor two, by
missing dust-obscured massive star forming systems that can now be
detected at sub-mm wavelengths.  The preliminary redshifts assigned to
the detected sub-mm sources would imply a rising co-moving star
formation rate to redshifts of 3 at least (Blain et al.~\cite{blain99};
see Trentham et al.~[\cite{trentham99}] for a deviating interpretation).
This excess star formation seems to occur in galaxies with star
formation rates between 10$^2$ M$_\odot$ year$^{-1}$ and 10$^3$
M$_\odot$ year$^{-1}$ (Lilly et al.~\cite{lilly99}), i.e., up to ten
times that observed for local ultraluminous IRAS galaxies, and over ten
times higher than that inferred for typical high-$z$ galaxies seen in
optical studies.

Unfortunately, our understanding of the faint sub-mm source population
remains poor, principally due to the limited number of reliable
identifications of sources in the optical or near-infrared (near-IR).
Moreover, for those sour\-ces thus far identified, many are associated
with faint optical and near-IR counterparts, which have typical $K$
magnitudes $\approx$20 to 21, and are occasionally very red, $I-K>6$
(Smail et al.~\cite{smail99}; Barger et al.~\cite{barger99a}; Dey et
al.~\cite{dey99}).  Hence, reliable optical and near-IR source
identifications require accurate source positions.  The sub-mm
observations described above were made with the Submillimeter Common
User Bolometer Array (SCUBA) at the James-Clerk-Maxwell Telescope
(JCMT), whose $14.\!''5$ FWHM beam at 850~$\mu$m is too large to allow
for proper source identification due to confusion in near-IR images
(Downes et al.~\cite{downes99}; Barger et al.~\cite{barger99b}; Smail et
al.~\cite{smail99}).

In order to increase the number of sub-mm/mm background sources with
accurate positions, we have begun an extensive program using the IRAM
Plateau de Bure radio Interferometer (PdBI) at 1.25 mm and the Very
Large Array (VLA) at 20~cm wavelength.  The PdBI was used to obtain
accurate positions for the strongest source each in the Hubble and
Canada-UK deep SCUBA fields detected by Hughes et al.~(\cite{hughes98})
and Eales et al.~(\cite{eales99}), respectively.  Both sources were
detected, allowing the determination of their positions to a fraction of
one arcsecond.  The detection and possible identification of the
HDF850.1 source was presented by Downes et al.~(\cite{downes99}), who
find that the source falls in between a faint arc-like structure at a
redshift between 1.7 and 3, and an elliptical galaxy at redshift 1.1,
thus not permitting an unambiguous identification.  The case is further
complicated by the possible gravitational lensing of the sub-mm source
by the elliptical galaxy, and by the lack of a significant radio
counterpart.  Although interesting in its complexity, the lack of a
clear identification of the HDF source in the optical and near-IR limits
its contribution to clarify the general nature of faint sub-mm sources.

In the following, we first present our PdBI observation of the brightest
source detected in the Canada-UK Deep Submillimeter Survey carried out
at 450\,$\mu$m and 850\,$\mu$m with SCUBA at the JCMT (Eales et
al.~\cite{eales99}) in fields used for the Canada-France Redshift Survey
(Lilly et al.~\cite{lilly95}).

Furthermore, we describe our program at the IRAM 30~m telescope, which
entails wide field surveys of regions that have been observed to very
faint levels with the VLA.  To date, all blank field searches for sub-mm
sources were conducted with only one instrument, SCUBA at the JCMT.  The
Max-Planck Millimeter Bolometer array (``MAMBO'', Kreysa et
al.~\cite{kreysa99}) at the IRAM 30\,m provides comparable sensitivity,
even taking into consideration the sharply rising spectrum
of the thermal sources.  The high resolution, sensitive VLA images are
fundamental to source identification, providing sub-arcsecond source
positions and a rough indication of source redshift when compared to the
mm flux density (Carilli \& Yun \cite{CY99}; Blain \cite{blain99}; Dunne
et al.~\cite{dunne00}).  In this paper we present the first two source
identifications from this program, including deep near-IR and optical
imaging.

\section{Observations and Results}
\label{se:obs}

\subsection{Previous observation of CFRS14A}

With an 850~$\mu$m flux density of $8.8\pm1.1$\,mJy, CFRS14A (Table
\ref{table1}) is the strongest sub-mm source detected by Eales et
al.~(\cite{eales99}) during the Canada-UK Deep Submillimeter Survey.  A
sensitive VLA 5\,GHz survey of the field by Fomalont et
al.~(\cite{fomalont91}) reveals the weak radio source 15V18 nearly
coincident with the SCUBA source.  The radio source has a flux density
of $44.0\pm4.1~\mu$Jy and may be extended by $\sim 1.5$\,arcsec.  Lilly
et al.~(\cite{lilly99}) obtained deep ground based and HST images at
the VLA position and found an optical/near-IR counterpart with $U_{\rm
  AB}=26.5$, $V_{\rm AB}=25.5$, $I_{\rm AB}=24.1$, and $K_{\rm
  AB}=20.8$, the colors of which suggest the galaxy to be at a redshift
$z\approx 2$.  Hammer et al.~(\cite{hammer95}) detected the source at
the same $K$ magnitude, and while they suggested it might be
elongated by $\approx 2''$, Lilly et al.\ find it more compact, $<1''$,
but ``not completely symmetrical.''

\begin{table*}[]
\caption{Source positions.\label{table1}}
\begin{tabular}[t]{llllll}\hline \hline
CFRS14A & SCUBA 850\,$\mu$m & $14^{\rm h}\,17^{\rm m}\,40.3^{\rm s}$ &
$52^\circ\,29'\,08''$ & 
2000 & Lilly et al.~\cite{lilly99} \\
15V18   & VLA 5\,GHz        & $14^{\rm h}\,17^{\rm m}\,40.21^{\rm s}\pm 
0.03^{\rm s}$ & $52^\circ\,29'\,06.\!''5\pm 0.\!''15$ & 
2000  & Fomalont et al.~\cite{fomalont91} \\ 
%
%
        & $U~ V~ I~ K$          & $14^{\rm h}\,17^{\rm m}\,40.33^{\rm s}$ &
$52^\circ\,29'\,5.\!''9$ & 
J2000 &  Lilly et al.~\cite{lilly99} \\
      & PdBI 1.25 mm & $14^{\rm h}\,17^{\rm m}\,40.30^{\rm s}\pm 0.04^{\rm s}$ &
 $52^\circ\,29'\,06.\!''8\pm 0.\!''35$ & J2000 & this work \\
\hline
J154127+6616 & VLA 1.4\,GHz & $15^{\rm h}\,41^{\rm m}\,27.281^{\rm s}\pm 0.013^{\rm
  s}$ & $66^\circ\,16'\,17.\!''00\pm 0.\!''06$ & 
J2000 & this work \\
J154127+6615 & VLA 1.4\,GHz & $15^{\rm h}\,41^{\rm m}\,26.901^{\rm s}\pm
  0.009^{\rm s}$ &  $66^\circ\,14'\,37.\!''29\pm 0.\!''07$ &
J2000 & this work \\ 
\hline
\hline
\end{tabular}
\end{table*}

\subsection{Interferometric observation of CFRS14A}

\begin{figure}[tb]
\epsfig{file=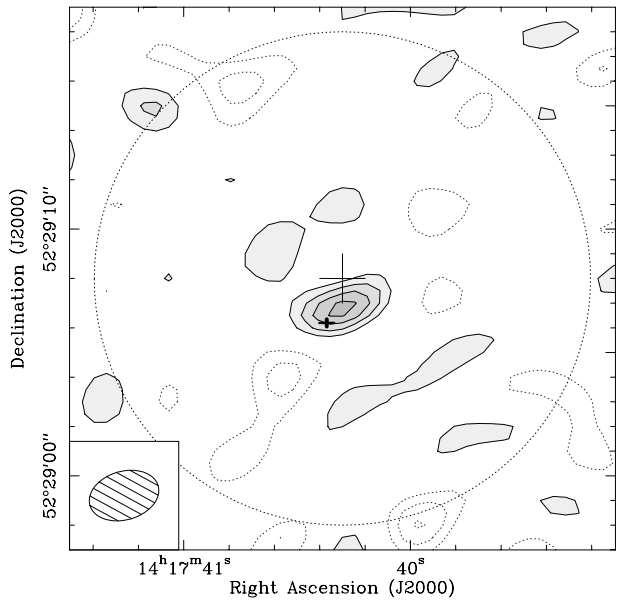,width=8.cm,angle=-90,
clip=,bbllx=50pt,bblly=50pt,bburx=550pt,bbury=600pt}
\caption{
  CFRS14A observed with the IRAM Plateau de Bure Interferometer in
  D-array at 240\,GHz (1.25~mm). Contours are spaced by 0.4~mJy/beam
  (1\,$\sigma$) starting at 1.5\,$\sigma$.  The clean beam (lower left
  corner) is $2.\!''9\times 2.\!''0$ at a position angle of $-74^\circ$.
  The position of the SCUBA 850~$\mu$m source is marked by a large
  cross, that of the Fomalont et al.~ 5\,GHz source 15V18 by a small
  cross. The cross arms reflect the 1\,$\sigma$ positional uncertainty.
\label{fig:fig1}
}
\end{figure}

We observed CFRS14A with the IRAM Plateau de Bure interferometer
(Guilloteau et al.~\cite{guilloteau92}) on 1998 November 25, 26 and 28.
The five antenna array was used only in its most compact configuration,
providing baselines extending up to 80~meters. The total integration
time was 16~hours. The dual-channel receivers were tuned to 105~GHz LSB
and 240~GHz USB. At 1.25~mm, data were taken in upper and lower
sidebands.  The correlator covered $320$~MHz at
105~GHz and $2\times500$~MHz at 240~GHz.

Good weather conditions provided typical SSB system temperatures of
200~K at 105~GHz and 300~K at 240~GHz and rms phase noise $\le 35^\circ$
at both frequencies. Flux densities were derived from observations of
MWC349 (adopted flux densities were 1.06~Jy at 105~GHz and 1.75~Jy at
240~GHz) and/or CRL618 (1.55~Jy at 105~GHz and 2.00~Jy at 240~GHz).
Temporal fluctuations of the phase and amplitude were calibrated by
frequent observations of the nearby quasars $1637+574$ and $1418+546$.
The final flux density accuracy at 1.25~mm is estimated to be $\sim
20\%$. Images were produced applying natural weighting and deconvolved using
CLEAN.

No source was detected at 105~GHz, to a $3\sigma$ limit of 0.6~mJy.  The
240~GHz data reveal a $2.0\pm 0.4$~mJy source (Fig. \ref{fig:fig1}).
The $5\sigma$ detection precludes a meaningful determination of an
angular size (see Downes et al.~[\cite{downes99}] for the discussion of
the similar case of HDF850.1).

The emission peaks $1.\!''2$ south of the 850$\mu$m SCUBA source,
$0.\!''7$ north-west of the radio source, and $0.\!''9$ north of the near-IR
source. The astrometric errors are $\sim 0.\!''4$ and $0.\!''2$ along
the beam major and minor axis, respectively. With a signal-to-noise
ratio of 5, the statistical error adds $\approx 0.\!''25$, so that the
total positional uncertainty is approximately $0.\!''35$.  Within their
errors the sub-mm, mm, radio, and near-IR positions are consistent with one
single source.  Future high-resolution VLA observations might better
establish any offset between the radio and near-IR source position.

\subsection{Sources from the MAMBO millimeter survey}

\begin{figure}[h] 
\epsfig{file=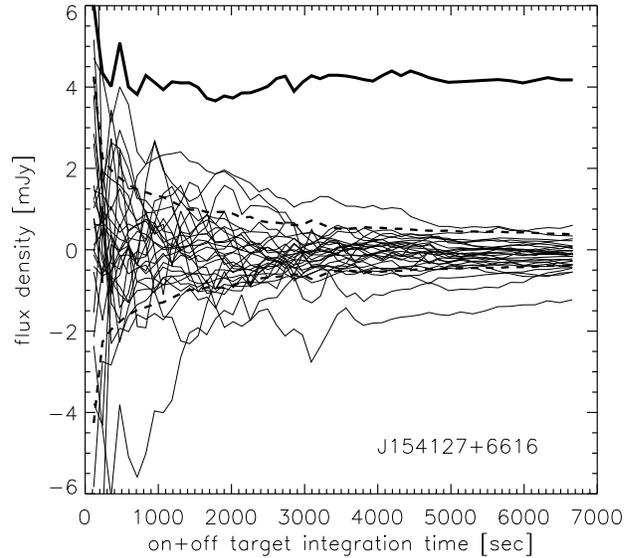,width=9.cm}
\caption{Time-averaged, cumulative
  1.2~mm signals of the individual bolometer channels as a function of
  on$+$off target integration time on J154127+6616, including all 1999
  and 2000 data.  The on-target central channel signal is plotted as a
  thick line, reaching a final value of $4.2\pm 0.3$\,mJy after 6664
  seconds.  The rms dispersion of the signal in the off-channels is
  plotted as broken lines, reaching $\pm 0.35$\,mJy.}
\label{fig:onof1}
\end{figure} 

\begin{figure}[h] 
\epsfig{file=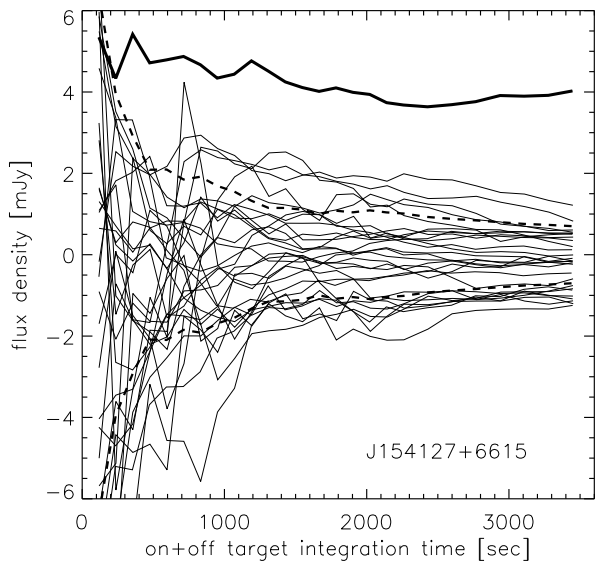,width=9cm}
\caption{Cumulative signal of individual bolometers for the on-off
  integration toward J154127+6615. The on-target channel signal (thick
  line) reaches $4.0\pm 0.5$\,mJy after 3446 seconds of on$+$off target
  integration. The signal dispersion (broken lines) in the off-channels
  reaches a final value of $\pm 0.7$\,mJy.}
\label{fig:onof2}
\end{figure} 

The Max-Planck Millimeter Bolometer 37-element array ``MAMBO''
(effective frequency 250\,GHz, bandwidth $\approx 80$\,GHz) at the IRAM
30~m telescope was used from February to April 1999 and in December 1999
to March 2000 to map three fields with a total area of over 300
arcmin$^2$ to 1\,$\sigma$ noise levels below 1~mJy (Bertoldi et al., in
prep.).

We here present two of the strongest mm sources found in our map toward
the $z=0.24$ cluster Abell 2125, of which we had previously obtained a
deep 1.4~GHz VLA image ($\rm rms \approx 7.5\mu Jy/beam$, Owen et al.,
in preparation). Both mm sources stood out in the MAMBO maps, and
coincided with VLA radio sources.  The VLA radio positions were then
targeted with deeper on-off bolometer integrations, the results of which
are displayed in Figs. \ref{fig:onof1} and \ref{fig:onof2}.

The on-off observations were conducted in standard chop-nod mode, with
individual scans of 3 minutes, divided in 12 or 16 subscans of which
each yields 10 seconds on$+$off source exposure.  The secondary mirror
was chopped by about 50$''$ in azimuth at 2 Hz, and the telescope was
nodded by the same distance after each subscan. The pointing accuracy is
typically better than 2\,arcsec.  The data were analyzed with the MOPSI
package (Zylka \cite{zylka98}). ``Skynoise'', i.e., the rapid variation
of the sky signal, was subtracted from each channel by subtracting the
correlated signal of the surrounding six channels.  The flux was
calibrated with observations of Mars, Uranus, and Ceres, yielding 4900
counts/Jansky for the early 1999 data, and 12,500 counts/Jansky in early
2000, which we estimate to be accurate within 10\%.

The observations of J154127+6616 were performed on five different days
under good atmospheric conditions.  The 52 scans add to 6795 seconds
on$+$off source integration. 
 The integrations of early 1999 yield almost
identical results to those from early 2000: $4.18\pm 0.44$\,mJy in 3805
sec for 1999, $4.16\pm 0.42$\,mJy in 2858 sec for 2000. The merged data
(Fig. \ref{fig:onof1}) yield $4.17\pm 0.31$\,mJy. The quoted error is the
integrated noise level of the central on-target channel. The signals
of the off-target channels have a similar dispersion.

The observations of J154127+6615 were performed on three different days,
16 February 1999, 9 April 1999, and 24 March 2000, and add to 3446
seconds on$+$off target integration time. The individual observations
yield less consistent results than for J154127+6616, possibly due to
pointing problems or unstable atmospheric conditions. The individual
observations yield source signals of $4.4\pm 1.1$\,mJy in 1069 sec,
$3.0\pm 0.8$\,mJy in 1190 sec, and $4.8\pm 0.6$\,mJy in 1188 sec, on the
respective dates.  The merged data (Fig. \ref{fig:onof2}) yield $4.0\pm
0.5$\,mJy.  The final dispersion in the off target channels is 0.7\,mJy.
The actual source flux may be somewhat above 4\,mJy, considering that the
most recent detection gave a higher flux.

The positions of the 1.4\,GHz VLA radio sources identified with the mm
sources are given in Table 1.  The 1.4\,GHz size of J154127+6616 is
$<0.\!''9$, and its integrated flux is $67\pm 13~\mu$Jy.  The more
southern source, J154127+6615, has a 1.4\,GHz size $<1.\!''3$ and an
integrated flux density of $81\pm 13~\mu$Jy.

\begin{figure}[tb] 
\center
\epsfig{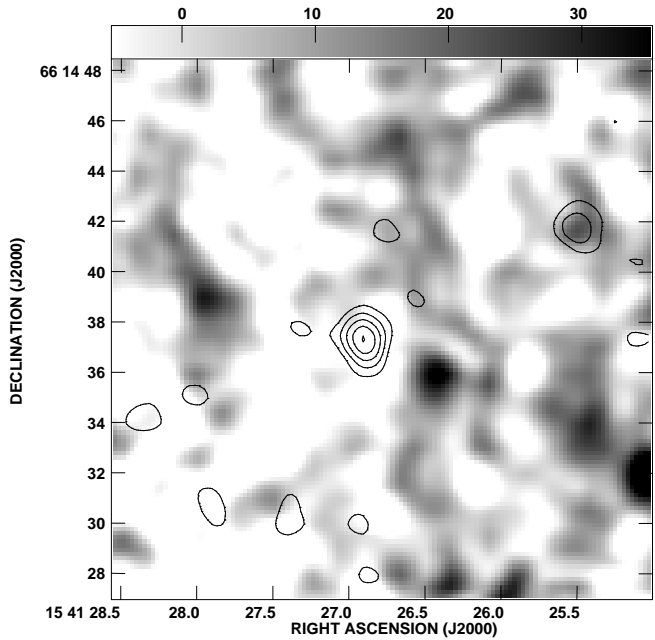}
\caption{APO $J$-band image toward the mm and VLA 1.4\,GHz source
  J154127+6615. The VLA 1.4\,GHz flux density is shown as contours with
  levels 15, 30, 45, 60\,$\mu$Jy/beam, J154127+6615 being near the image
  center.  No $J$ counterpart is found to 23.5\,mag. The source is also
  not see in the $R$-band to 24.5\,mag and in $K$ to 20.0\,mag.}
\label{fig:J-VLA-south}
\end{figure} 

\begin{figure}[tb] 
\center
\epsfig{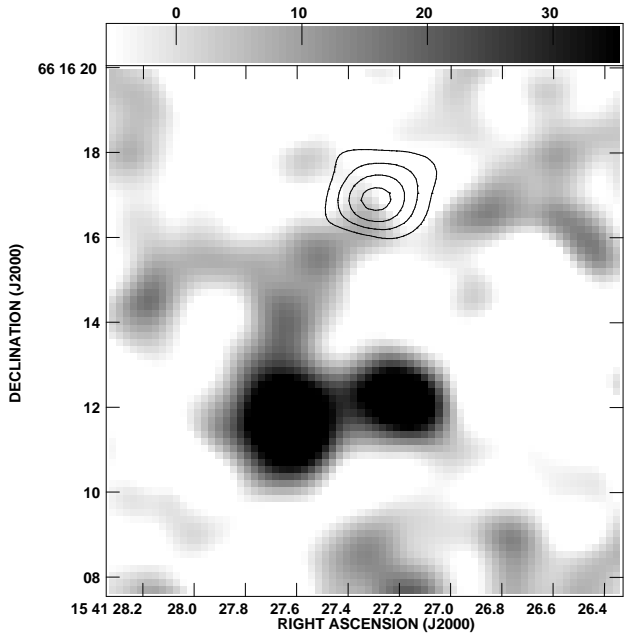}
\caption{APO $J$-band image toward J154127+6616 
  with VLA 1.4\,GHz intensity as contours (same levels as in
  Fig.\ref{fig:J-VLA-south}). The source seen at 1.2\,mm, 1.4 GHz, and in
  the $K$-band (see Fig.\ref{fig:K-VLA-north}) 
  is not visible in $J$ to 23.5\,mag.  However, the faint
  arc connecting this source with a neighboring $K$ source is apparent
  in both the $J$ and $K$ images (see Fig. \ref{fig:K-VLA-north}).}
\label{fig:J-VLA-north}
\end{figure} 

\begin{figure}[tb] 
\center
\epsfig{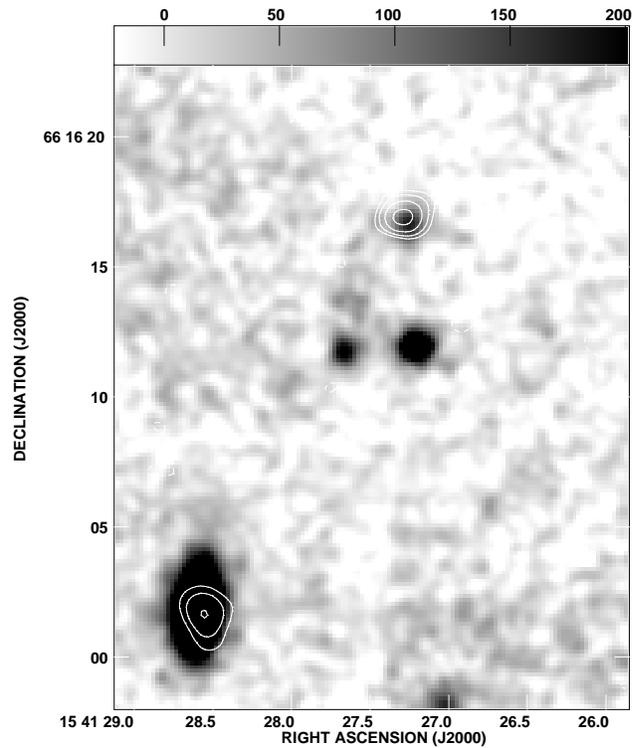}
\caption{Keck $K$-band image toward J154127+6616, showing the 1.4\,GHz 
  (contour levels as in Fig.\ref{fig:J-VLA-south}) source in the upper
  part of the image to coincide with a $\sim 20.5$\,mag $K$ source.  An
  unrelated bright radio and near-IR source is seen in the south-east
  corner of the image.}
\label{fig:K-VLA-north}
\end{figure} 

\subsubsection{$R$-band}

An $R$-band image including both mm/radio sources was taken with the
KPNO 0.9 m telescope.  It does not show any counterpart to either source
brigher than $R=24.5$\,mag.

\subsubsection{$J$-band}

A $J$-band image was obtained at the Apache Point Observatory (APO) with
the ARC 3.5 m telescope on 1 July 1999. The detector was the GRIM II
256$\times$256 NICMOS array in F/5 mode with an image scale of
0.48\,arcsec pixel$^{-1}$. Sky conditions were photometric and the
seeing at $J$ was $\sim 0.\!''8$. Zero point calibration was determined
from observations of ARNICA IR Standard stars (Hunt et
al.~\cite{hunt98}).

The two MAMBO 1.2\,mm identifications have a separation of 100\,arcsec
on the sky. As the field-of-view of the array was 120\,arcsec, we were
constrained to dither the field in E--W, keeping both positions in the
FOV at all times. We exposed for 60\,sec at each of the 5 dithered
positions separated by 10\,arcsec, for a total exposure time of 91
minutes.

The absolute positional errors of the images are $\approx 0.\!''2$
($1\sigma$). The coordinate system was derived from the radio frame
($\approx 0.\!''05$ error), using the optical identifications to fix the
solution on the $R$-band image. The $J$-band image was registered using
objects from the $R$ image which also appear in the $J$ image. The same
was done with the KPNO and Keck $K$-band images.

The $J$-band image (Figs.~\ref{fig:J-VLA-north} and
\ref{fig:J-VLA-south}) does not show any counterpart to both mm/radio
sources brigher than $J=$ 23.5\,mag.  However, a $J=22.5$\,mag object
appears about 3\,arcsec away from J154127+6615, which we believe to be
unrelated to the radio source, although the possibility that it is the
mm counterpart is not excluded; we did not target this source with
on-off bolometer observations.

\subsubsection{$K$-band}

A $K$-band image obtained with the Ohio-State NOAO Infrared Spectrometer
on the 2.1\,m telescope at the Kitt Peak National Observatory does not
show any counterpart to either J154127+6615 or J154127+6616 brighter
than $K=20$\,mag ($3\,\sigma$ limit in a 2\,arcsec aperture).

A deep $K$-band image obtained using the Near Infrared Camera (NIRC;
Matthews \& Soifer \cite{matthews94}) on the Keck\,I 10\,m telescope
reveals a $K=20.5$\,mag counterpart to J154127+6616, and two more, $\sim
20.5$~mag and 19.7\,mag, sources $\sim 4''$ south of it
(Fig.~\ref{fig:K-VLA-north}).  The Vega-based magnitudes were measured
in a 3$''$ aperture. The identification of the 1.2\,mm source with the
northern of the three $K$ sources is most likely, because we did obtain
MAMBO on-off observations pointed about 4\,arcsec south and south-east
of the VLA source position, which did not produce a signal at a
comparable level.

The $K$-band image shows a faint arc connecting the mm/radio counterpart
with the $K$ source 4$''$ south-west of it.  This arc is also seen in
the $J$-band image (Fig.~\ref{fig:J-VLA-north}), whereas no $J$-band
counterpart appears at the position of the radio source. If this arc is
real, it could suggest an association of the northern and southern
sources in this system.  Perhaps we observe an interacting group of
galaxies, in which the northern source contains a dust-obscured
starburst.  However, all three sources have very different colors, with
the southern source colors being more typical of low redshift sources.
This may be a hint against their association with the very red
$K$/mm/radio source.

The possible arc is similar to structures seen toward several
high-redshift sources, e.g., the $z = 4.7$ dust emitting quasar BR
1202$-$0725 (Ohta et al.~2000). Arcs and multiple sources are indicative
of lensing. However, the color differences between the three sources and
the arc are significant, suggesting that they are not likely to be
lensed images of one and the same source.  The foreground cluster A2125
may magnify both mm source intensities by gravitational lensing, but
since both sources are relatively far from the cluster center, the
effect should be small.


\section{Discussion}

The faint sub-mm source in the CFH field, and the two sources in the
A2125 field, follow a pattern that seems typical for many of the
detected sub-mm sources. They have 10 to 100\,$\mu$Jy radio continuum
counter-parts at 20~cm, they have very faint near-IR counter-parts, and
they are mostly very red.

\subsection{Redshift estimates}

CFRS14A has a 350\,GHz flux density of 8.8\,mJy and a 240\,GHz flux density
of 2.0\,mJy.  The 350\,GHz to 240\,GHz flux ratio of CFRS14A implies a
steep sub-mm spectral index,
\begin{equation}
\alpha_{\rm submm} = { \log[(8.8\pm 1.1)/(2.0\pm 0.4)]
\over \log(350/240)} ~=~ 3.5\pm 0.5~.
\end{equation}
Since such a value is typical for nearby starburst galaxies such as Arp
220 ($\alpha_{\rm submm}\approx 3.0$) or M82 ($\alpha_{\rm submm}\approx
3.5$), the spectral index of CFRS14A does not indicate a flattening of
the spectrum which would be found when the dust emission peak is
redshifted to near 850\,$\mu$m. This indicates that the object may not
be at very high redshift, i.e., at $z<3$, which is consistent with the
photometric redshift of $\approx 2$ derived from the observed
optical/near-IR colors.

Carilli and Yun (\cite{CY99}, \cite{CY00}; CY) have shown that the
radio--to--sub-mm spectral index provides a rough indication of redshift
for star forming galaxies, based on the tight radio--to--far infrared
correlation for star forming galaxies found by Condon (\cite{condon92};
see also Dunne et al.~\cite{dunne00}; DCE). This correlation however
assumes a narrow range of dust temperatures, because a high-$z$ object
with warm dust would show the same spectral index as one with cool dust
at low $z$ (Blain \cite{blain99}).  Redshift estimates based on the
radio--to--sub-mm spectral index are thus to be taken with caution.

CFRS14A has a radio flux density at 5\,GHz of 44\,$\mu$Jy.  Assuming a
radio spectral index of $-0.8$, typical for star forming galaxies
(Condon \cite{condon92}), implies an expected 1.4 GHz flux density of
120\,$\mu$Jy.  The radio--to--sub-mm spectral index is then 0.78, which
implies a most likely redshift of 1.9, with a reasonable lower limit of
$z >$ 1.3 using the (revised) mean-galaxy CY model based on 17 low
redshift star forming galaxies. The DCE model based on 104 low redshift
galaxies yields a most likely redshift of 1.6.  Both redshifts are,
within the errors, consistent with the photometric redshift estimate.

The two sources from the A2125 field have 250\,GHz flux densities of
about 4\,mJy.  The 1.4\,GHz flux densities are around 75\,$\mu$Jy,
leading to a spectral index between 1.4 and 250\,GHz of 0.77.

Using the observed galaxy spectra from the revised CY model, a spectral
index of 0.77 between 1.4 and 250\,GHz implies a most likely redshift of
2.5, with a reasonable lower limit (84$\%$ confidence level) of $z > 2$.
Using the DCE models leads to a most likely redshift of 2.2 (see
discussion below).

\subsection{Redshift distribution}

\begin{figure}[htb] 
\center
\epsfig{file=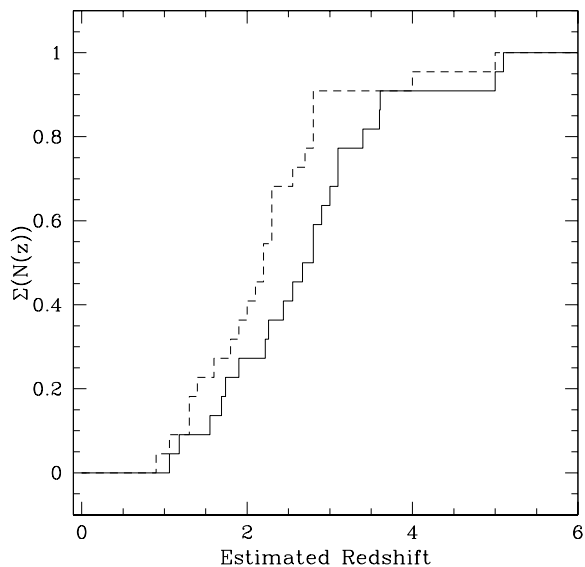,width=8cm}
\caption{
  The cumulative redshift distribution for faint sub-mm sources derived
  using $\alpha_{1.4}^{350}$ -- $z$ models. The distribution includes 22
  sources, as taken from this paper, from Downes et
  al.~(\cite{downes99}), Smail et al.~(\cite{smail00}), Barger et
  al.~(\cite{barger99b}), and Ivison et al.~(\cite{ivison00}).  The
  solid line is the distribution predicted using the (revised) $z_{\rm
    mean}$ model of Carilli and Yun (\cite{CY00}).  The dashed line is
  the distribution predicted using the model of Dunne et
  al.~(\cite{dunne00}).  Note that 7 of the sources in the Smail et
  al.~(\cite{smail00}) sample have only lower limits to the values of
  $\alpha_{1.4}^{350}$, so these curves should be considered strictly
  lower limits to the true distributions.  For four of the galaxies
  reliable spectroscopic redshifts are available, of which three agree
  with that derived from the CY model within 0.1, whereas the fourth
  disagrees, but is most likely a radio-loud AGN.  }
\label{fig:Zdist}
\end{figure}

We can use the observed or implied values of $\alpha_{1.4}^{250}$ for
the 3 sources discussed here, plus the 14 field galaxies in the Smail et
al.~(2000) sample, plus HDF850.1 and the three other sub-mm selected
sources in the HDF and the HDF-FF with sensitive radio limits (Downes et
al.~\cite{downes99}; Barger et al.~\cite{barger00}), plus a source
found by Ivison et al.~(\cite{ivison00}) toward Abell 1835, to derive
the redshift distribution of the faint sub-mm sources. For sources
observed at 350\,GHz we use the revised CY and the DCE models for the
$\alpha_{1.4}^{350}$ -- $z$ relationship.  The cumulative redshift
distribution for these 22 sources is shown in Figure \ref{fig:Zdist}.

The CY model leads to a median redshift of 2.8, while the DCE model
leads to a median redshift of 2.2.  For 7 of these galaxies only lower
limits to $\alpha_{1.4}^{350}$ are available, which leads to lower
limits to the possible redshifts. Most of these limits are in the range
of $z = 2$ to 3, leading to a steep rise in the redshift distribution in
this range.  It is likely that the distribution will become more gradual
when more sensitive radio observations are made of these sources.  It is
also important to keep in mind that these models are relying on the idea
that the radio--to--sub-mm correlation for star forming galaxies is
independent of redshift.

Part of the offset between the CY and DCE redshift estimates may be due
to the fact that the majority of the sources used by CY are in the upper
half of the luminosity distribution of galaxies in the DCE sample.  CY
and DCE find a systematic trend for decreasing $\alpha_{1.4}^{350}$ with
increasing luminosity.  In either case, these results strengthen the
primary conclusions of Smail et al.~(2000) that the majority of the
faint sub-mm sources are likely to be at $z \ge 2$, and that there is no
prominent low-$z$ tail in the distribution.  

The distribution of CY redshift estimates is very close to that
predicted from the fraction of uncollapsed $10^{12} \rm M_\odot$
structures in a standard cold dark matter Press-Schechter formalism with
a bias factor of 2 (Peebles \cite{peebles93}).  This agreement may be
fortuitous, but it is remarkable.

\begin{acknowledgements}
  We thank R.\,Lemke and B.\,Weferling for their support during the
  MAMBO observations, and the anonymous referee for thoughtful comments
  on the manuscript. This work is based on observations carried out with
  the VLA, the IRAM 30\,m and Plateau de Bure, Keck, Apache Point, and
  KPNO telescopes.  IRAM is supported by INSU/CNRS (France), MPG
  (Germany) and IGN (Spain).  We are thankful to the IRAM staff,
  especially U.\,Lisenfeld, A.\,Sievers and R.\,Neri, for their support
  with the observations and data reduction.  The VLA is a facility of
  the National Radio Astronomy Observatory (NRAO), which is operated by
  Associated Universities, Inc.\ under a cooperative agreement with the
  National Science Foundation.  Kitt Peak National Observatory is part
  of the National Optical Astronomy Observatories, which is a facility
  of the National Science Foundation operated under cooperative
  agreement by Associated Universities Inc. The W.M.\,Keck Observatory
  is a scientific partnership among the University of California, the
  California Institute of Technology, and the National Aeronautic \&
  Space Administration, and was made possible by the generous financial
  support of the W.~M.~Keck Foundation. The Apache Point Observatory is
  owned and operated by the Astrophysical Research Consortium.  C.C.
  acknowledges support from the Alexander von Humboldt Society.

\end{acknowledgements}

{ }
\end{document}